# AI based Scintillation Detector Calibration

Navaneeth P. R., Kajal Kumari, Mayank Goswami

*Department of Physics, Indian Institute of Technology Roorkee*

mayankfph@iitr.ac.in

***Abstract*—Data set generated from the scintillation detector is used to build a mathematical model based on three different algorithms: (a) Multiple Polynomial Regression (b) Support Vector Regression (c) Neural Network algorithm. Using visualizations and correlations it is found that Median of the data will give accurate results and average time has a major contribution in radiation measurement.**

***Keywords—Support Vector Regression, Multiple polynomial Regression, Neural Network, Spearman Correlation.***

## I. Introduction

Scintillation detectors are mainly used to detect ionizing radiations in a sample. Scintillation is the phenomenon of the production of flashes of light when radioactive radiations fall on a transparent material like Zinc Sulfide, Lithium Iodide, etc. These scintillations can be measured in terms of pulse current with the help of a photomultiplier tube (PMT) and an electrometer. The scintillation detectors have a high amplification value, so they are used for the detection of several radiations which other detectors cannot easily detect. Data from a certain detector consists of variables: time, amplifier value, LLD (Lower Limit Discrimination), High Voltage, and Counts. These variables were made to fit into various Machine Learning and Deep Learning algorithms, and results were analyzed for a certain detector. The visualization and correlations helped in further understanding the data generated in the experiment.

## II. Preprocessing the Dataset

### A. *Parsing Raw Data from given Detector Data*

The raw data was given in the form of text files, wherein each file had values of variables: LLD (Lower Limit Discrimination), High Voltage and Counts, the value for Amplifier value, and time was given in the file name. Each file had 30 runs of the experiment. A python script was used to parse the variable values from entire files given for a particular detector. Since the experiment was repeated 30 times at a particular setting, the mean of the variable 'Count' was calculated for each data point. The entire data points for the analysis were written into a CSV file.

### B. *Visualizing the variables*

Visualization of the variables was done using histograms, boxplots, and scatter plots to get a better insight into the data. From the histogram plots of the variables HighVoltage, Time, LLD, and Amplifier value, it is clear that the range in which each variable lies is different, so this ensures the need of scaling the dataset before feeding into the algorithms for modeling.

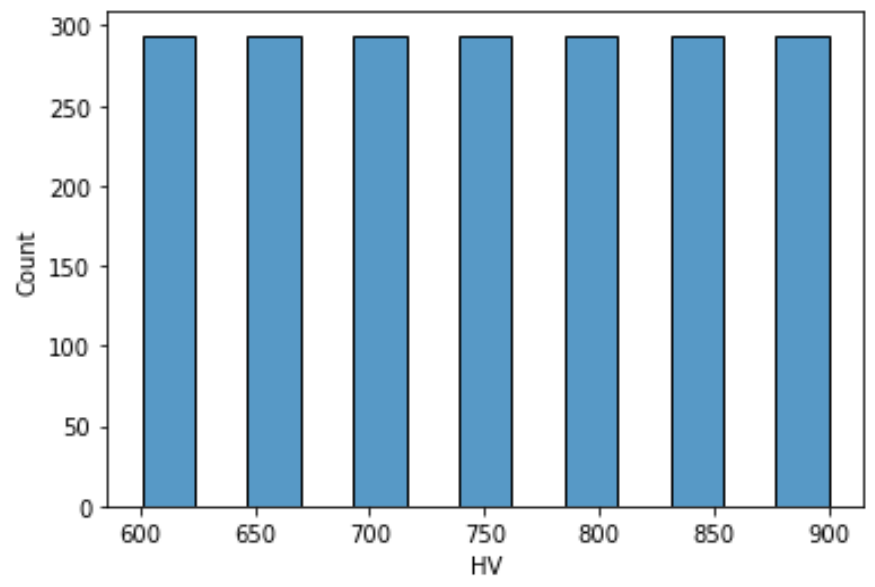

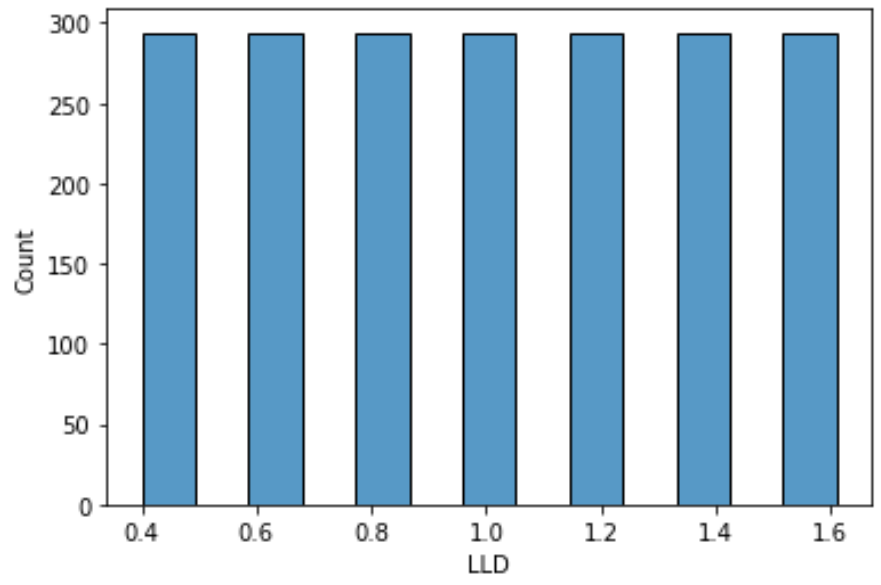

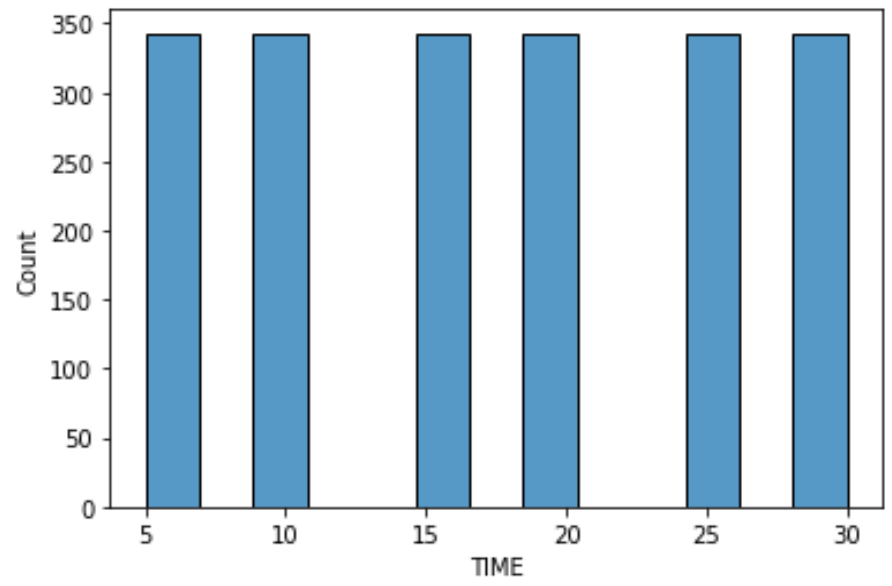

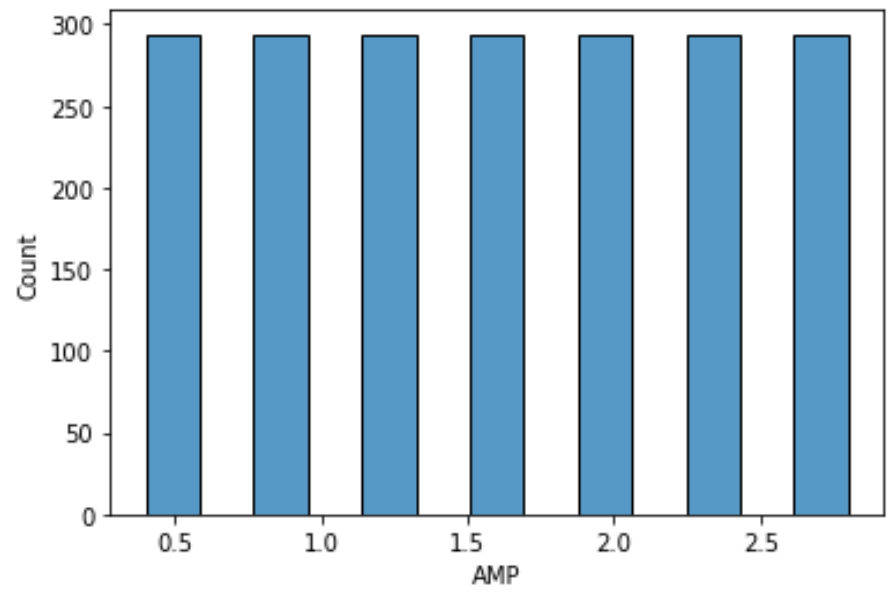

From the scatter plots of the variables, it is clear that the randomness of data points in the given dataset is low. Hence there might be a decrease in the accuracy of results predicted from the Machine Learning and Deep Learning Algorithms.

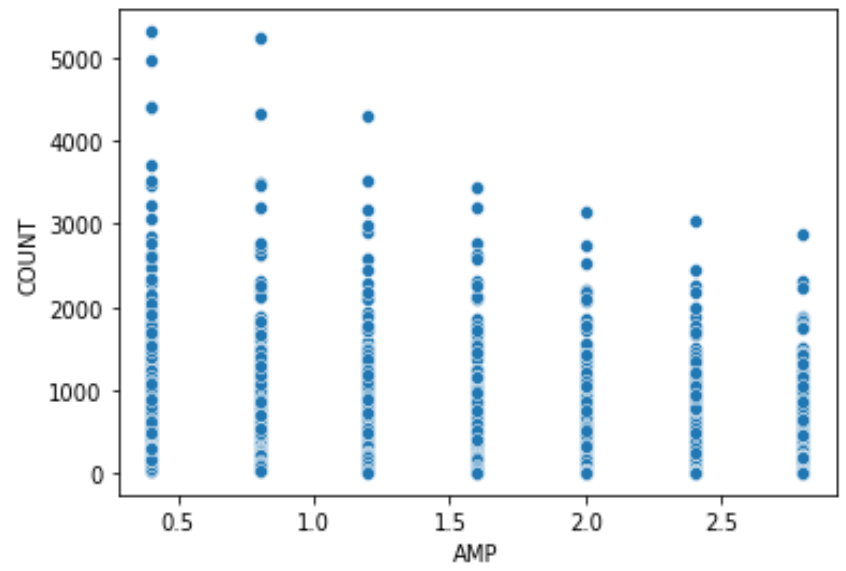

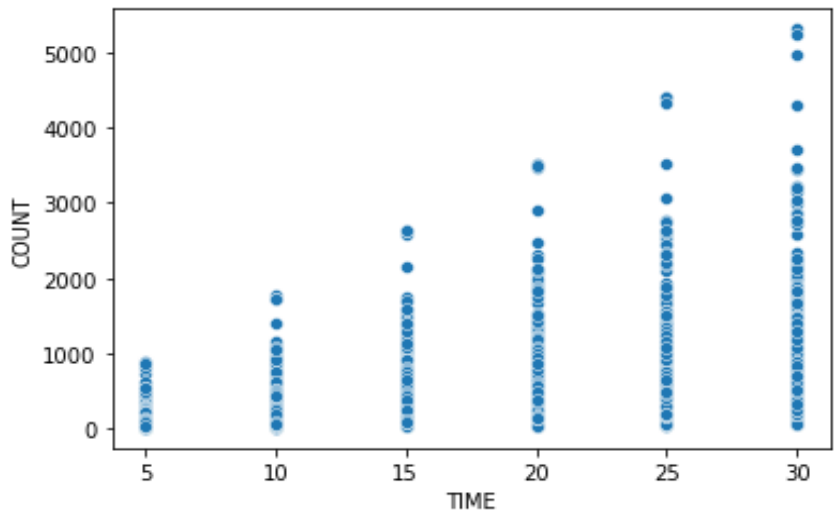

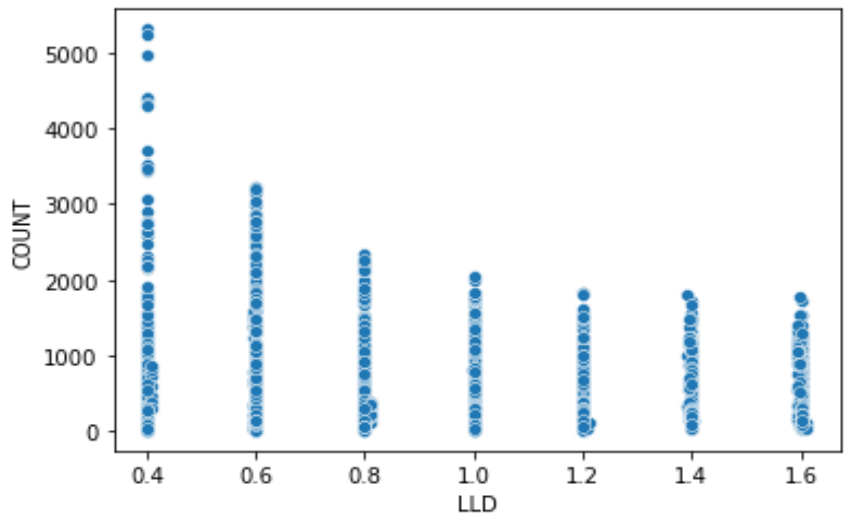

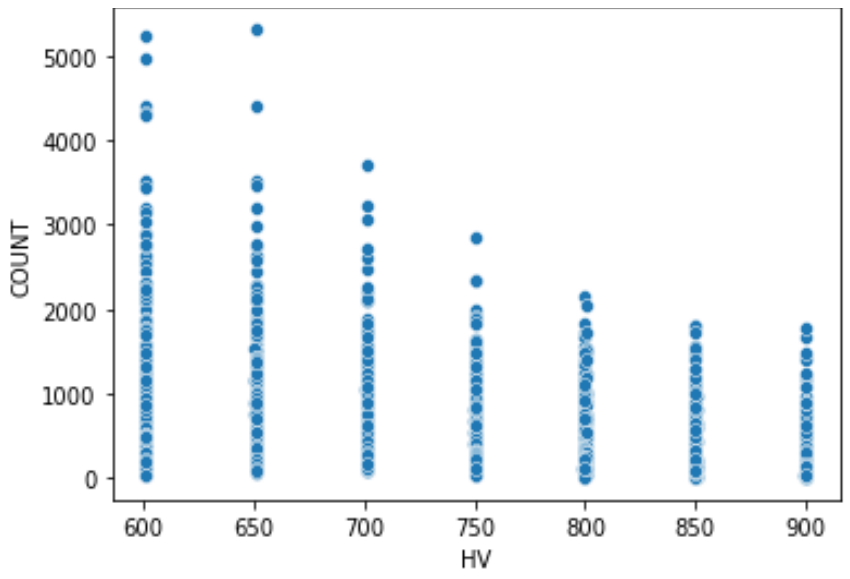

The Kernel Density Estimate plot of the Count variable shows that it follows a right-skewed distribution; hence the usage of Median as a measure of central tendency for analysis would yield more accurate results.

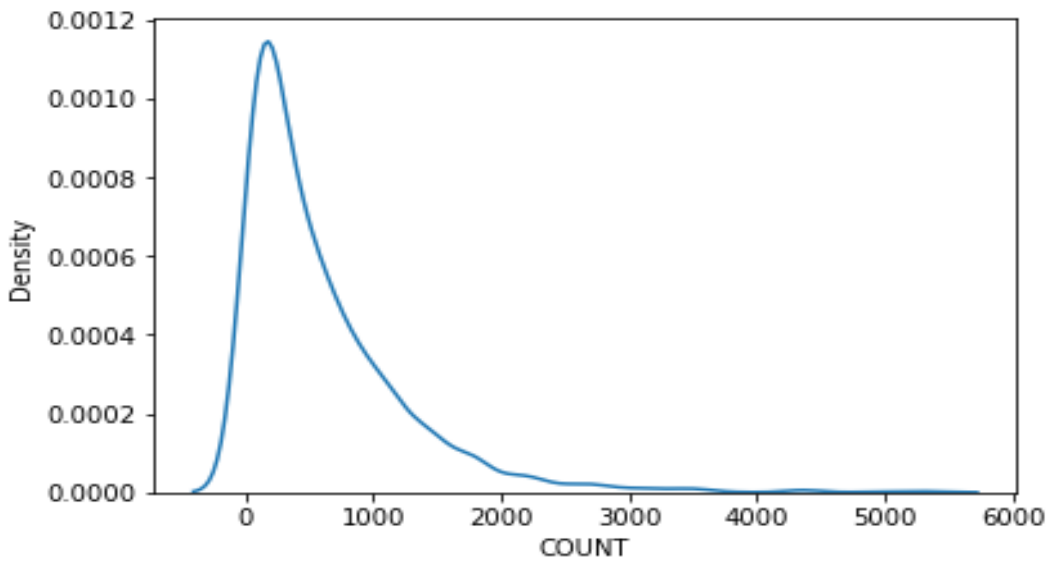

The box plot of the count variable here shows that there are a noticeable amount of outliers in the dependent variable, and this could affect the training time and accuracy of the algorithms used for training.

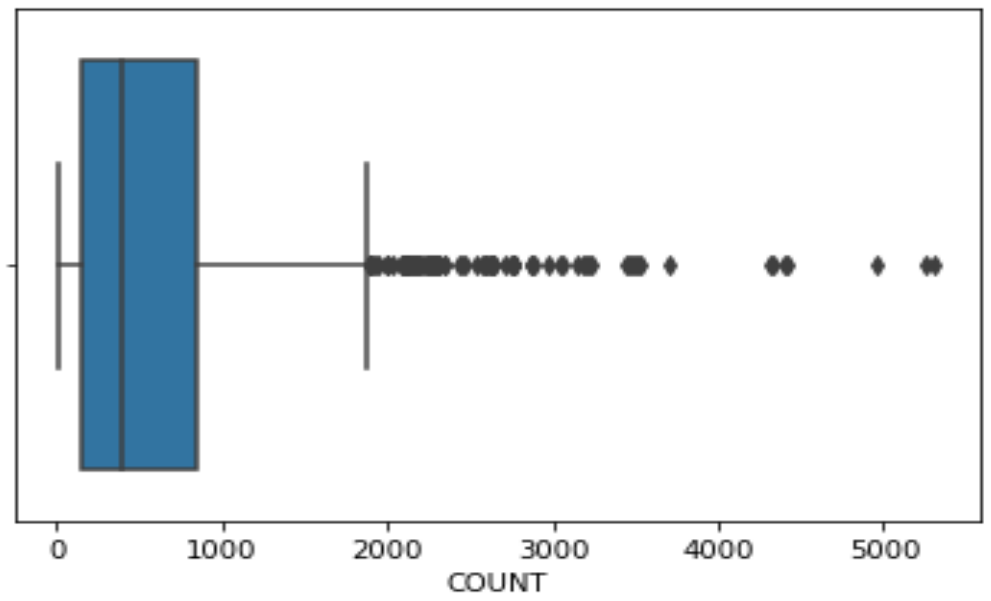

### C. *Scaling the Data*

All the variables in the dataset were scaled to have a mean of 0 and a standard deviation of 1. A particular variable in the dataset was scaled using the formulae given below.

$$x_{scaled} = \frac{x - mean(x)}{std.dev(x)}$$

Where mean and standard deviations of the variable were found using the given formulas.

$$mean(x) = \frac{sum(x)}{count(x)}$$

$$std.dev(x) = \left(\frac{sum((x - mean)^2)}{count(x)}\right)^{\frac{1}{2}}$$

## III. FINDING CORRELATIONS IN THE DATASET

Since the variables Time, Amplifier value, High voltage, and LLD exhibit a non-linear relationship with the count variable as mentioned in the article [1], a non-linear correlation was used to find the relationship between the independent and dependent variables. Spearman's Rank-Order Correlation was used for finding the correlation. Spearman correlation was preferred since it helps in determining the strength and direction of any monotonic relationship between variables that are non-linear.

| Independent Variable | Correlation Coeff. | p-value |
|---|---|---|
| Time | 0.5064 | 1.733e-134 |
| Amplifier Value | -0.326 | 3.66e-52 |
| High Voltage | -0.5005 | 6.224e-131 |
| LLD | -0.0118 | 0.5926 |

Here we are considering a null hypothesis that the variables are uncorrelated, and then the values of correlation coefficient and p-value were calculated between each independent and dependent variable (Counts).

From the above table, for variables with p-values less than 0.05, we can neglect our null hypothesis. Therefore, the variables Time, Amplifier Value, High Voltage have a greater correlation with the count variable. Among those variables, the time has a correlation coefficient of 0.5064 and has the lowest p-value, which suggests that time has a major impact on the determination of Counts. The results from the above correlations are in accordance with the non-linear equation given in the article.[1]

$$Counts = (-0.125) \times (LLD)^{-13.50} + (2.4) \times (Time)^{1.67} + (-3.09) \times (Amp)^{4.93} + (0.63) \times (H.V)^{1.09} + 3.34$$

## IV. Fitting Machine Learning and Deep Learning Models

Since there exists a non-linear relationship between the independent variables and the Count variable, mainly three algorithms that could accommodate non-linear data were chosen to model the Scintillation detector data.

### A. *Deep Learning Model*

Using a neural network to fit the dataset, which consists of 4 layers of neurons. The input layer had 4 neurons, the 2 hidden layers after the input layer had 5 neurons each, and the output layer had only 1 neuron. A sigmoid function was used as the activation function in all subsequent layers after the input layer.

A combination of the following optimizer and loss functions were found to be accurate: A gradient-based optimization technique RMSprop was used for optimization, and the mean absolute error was used as the loss function. The learning rate used was an exponentially decaying learning rate with the initial value of 0.05 and decay rate of 0.96.

The model was made to train with data in batches of 4, and a total of 100 epochs was performed. The model was evaluated to have a 45% accuracy in predicting the results. This was rather expected from the insights derived in the data preprocessing phase.

### B. *Multivariate Polynomial Regression Model*

The data was made to fit into a non-linear polynomial of maximum degree 20. Such a model was experimented with because of the flexible nature of the polynomials, their ability to fit a wide range of curvature. Necessary steps like Backpropagation and Elimination of variables based on p-values and correlations were not performed, taking into the fact that the variables were non-neglected in the article [1].

The algorithm was trained using 80% of the data points chosen randomly from the given dataset, and 20% was used for testing.A scatter plot was used to visualize the predictions and original data. From the plot, it is clear that the model is not able to make accurate predictions, mainly for data points below the range of 100.

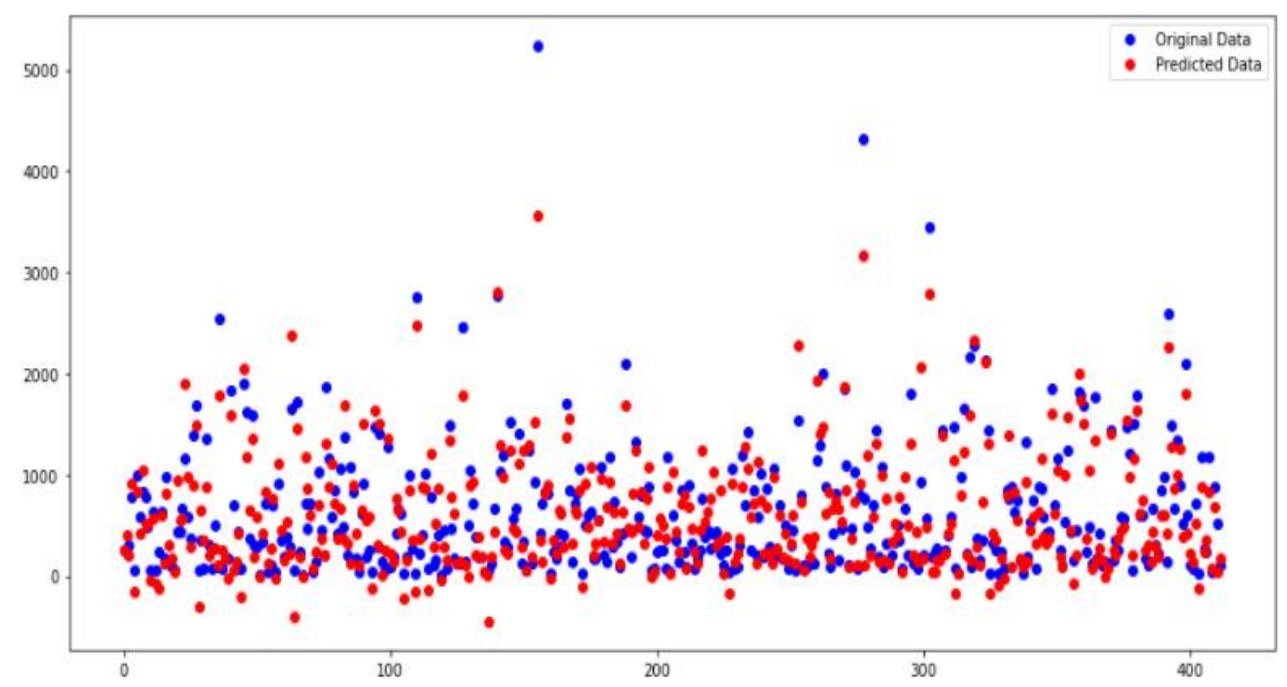

### C. *Support Vector Regression*

SVR model was chosen because it reformulates the problem into an optimization problem to find an error insensitive region or tube that best approximates the given continuous-valued non-linear function.

It also balances the model complexity with prediction error since both attempts to rigidly fit the data in the two methods discussed above had visible errors that cannot be neglected. Applying this model was a loose experiment of treating the above problem as an optimization problem rather than a regression problem. *A Gaussian- Radial Basis Function was used as the kernel since it gives a stable performance than using a polynomial kernel to map the points to higher-dimensional space.*

Training of the algorithm was done similarly to the multivariate polynomial regression algorithm. Scatter plots were used to visualize the predicted and original values of counts. From the plot, it is clear that the SVR model is able to accommodate errors more flexibly and predict more accurately than both of the above-mentioned methods. It is also visible that there is much overlap between the original and predicted data in the SVR model than the multivariate polynomial model.

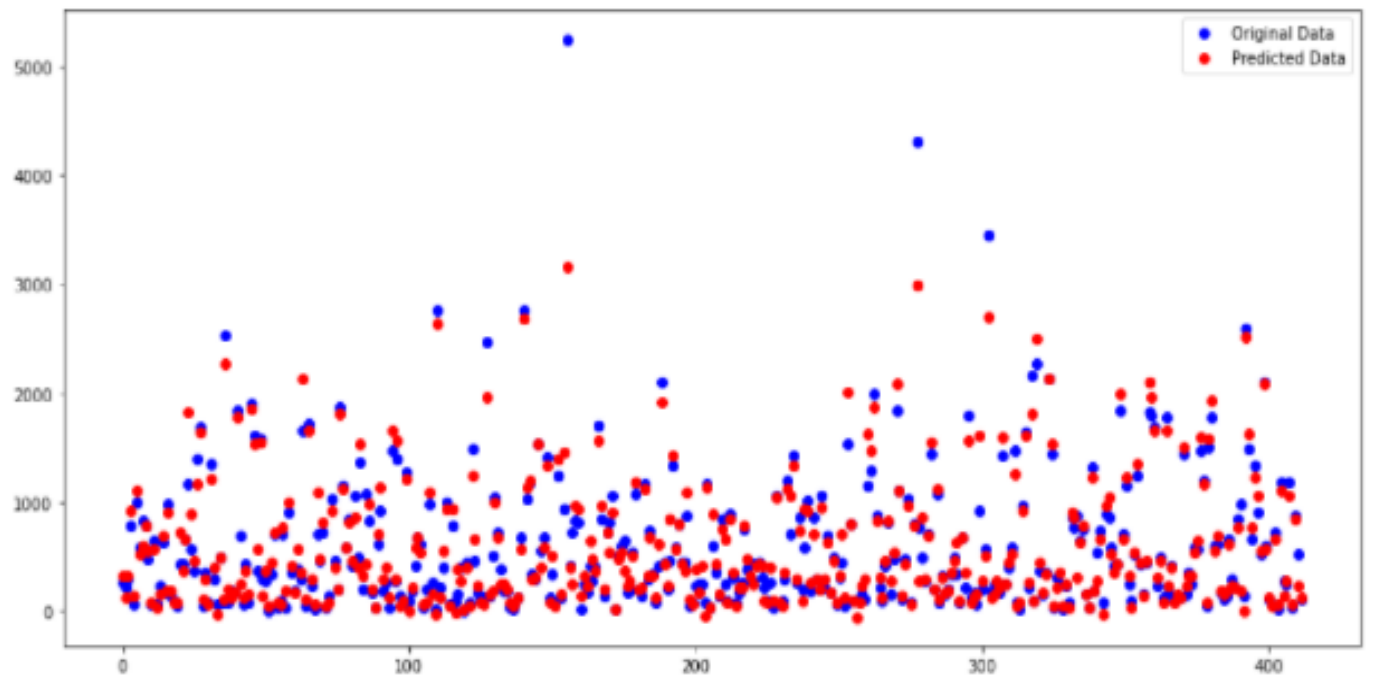

## Acknowledgment

The work is partially supported by the DST-SERB under Early Career scheme, Project number: ECR/2017/001432.